\documentclass[final]{raa}            % referee version: for submission

\usepackage{graphicx,times}             %for PS/EPS graphics inclusion, new
\usepackage{natbib}
\usepackage{amssymb,amsmath}
\usepackage{gensymb}
\usepackage{epstopdf}
\usepackage{physics}
\usepackage{multirow}
\usepackage{soul} 
\usepackage{float}
\bibpunct{(}{)}{;}{a}{}{,}

\usepackage{color}

\usepackage[pagebackref=true]{hyperref}
\hypersetup{
    colorlinks=true,
    linkcolor=blue,
    filecolor=magenta,
    urlcolor=blue,
    citecolor=blue,
    pdfpagemode=FullScreen,
    }

\newcommand{\MHz}{\ensuremath{\, {\rm MHz}}}

\begin{document}

\title{Beamforming in Interferometer Arrays with Cross-couplings}
%   \subtitle{I. Place Your Subtitle Here}

   \volnopage{Vol.0 (20xx) No.0, 000--000}      %%preserved for Editor. DOn't remove!
   \setcounter{page}{1}          %%starting page, preserved for Editor. DOn't remove!

   \author{Yingfeng Liu 
      \inst{1,2}
    \and Shijie Sun
      \inst{1,3}
    \and Kaifeng Yu
      \inst{4}
   \and Furen Deng
      \inst{1,2}
    \and Shifan Zuo
    \inst{1,3}
    \and Jixia Li
    \inst{1,3}
    \and Yougang Wang
    \inst{1,2,3}
    \and Fengquan Wu
      \inst{1,3*}
   \and Xuelei Chen
      \inst{1,2,3,\dag}
   }

   \institute{National Astronomical Observatories, Chinese Academy of Sciences,
             Beijing 100101, China; *{\it wufq@bao.ac.cn}, \dag{\it xuelei@cosmology.bao.ac.cn}\\
        \and
             University of Chinese Academy of Sciences Beijing 100049, China;\\
        \and
            Key Laboratory of Radio Astronomy and Technology, Chinese Academy of Science, Beijing 100101, China;\\
         \and 
            School of Physics and Astronomy, Sun Yat-Sen University, Zhuhai, 519082, China
      \vs\no
   {\small Received 20xx month day; accepted 20xx month day}}

\abstract{For an interferometric array, an image of the sky can be synthesized from interferometric visibilities, which are the cross-correlations of the received electric voltages of pairs of array elements. However, to search for transient targets such as the fast radio burst (FRB), it is more convenient to use the beam-forming technique, where the real-time voltage outputs of the array elements are used to generate data streams (beams) which are sensitive to a specific direction. This is usually achieved by a weighted sum of the array element voltages, with the complex weight adjusted so that all outputs have the same phase for that direction. Alternatively, beams can also be formed from the weighted sum of the short time averaged correlation (visibility) data. We shall call these two approaches the electric voltage beam forming (EBF) and cross-correlation beam forming (XBF), respectively.  All beams formed with the EBF can also be formed by the XBF method, but the latter can also generate beams which can not be generated by the former. We discuss the properties of these two kinds of beams, and the amount of computation required in each case. For an array with large number of elements, the XBF would require much more computation resource, although this is partly compensated by the fact that it allows integration over time. We study the impact of cross-coupling between array elements on the beamforming, first using a toy model, then for the case of the Tianlai Cylinder Pathfinder Array. In both cases, we find that the impact of the cross-coupling on the beam profile is relatively small. The understanding gained in this study is helpful in designing and understanding the beam-forming FRB digital backend for compact arrays such as the  Tianlai array. 
}

   \authorrunning{Yingfeng Liu et al. }            %author_head in even pages
   \titlerunning{Cross-Correlation Beamforming }  % title_head in odd pages

   \maketitle

%% Note: In the following text body of your manuscript, please note several differences from
%%       other major journals:
%% (1) \subsection{Please Capitalize the First Letter of Each Notional Word in Subsection Title}
%% (2) Please Capitalize the First Letter of Each Notional Word in all tables' captions

%
%________________________________________________ sections below
%\maketitle

\section{Introduction}
Interferometer arrays are widely used in radio astronomy. During its observation, the voltage output of each array element is amplified and cross-correlated with other elements, producing the so-called (interferometric) visibilities. Denoting the voltage of elements $a$ as $\mathcal{E}_a$, it is given by
\begin{equation}
\mathcal{E}_a = g_a \int e^{-2\pi i {\bf n}\cdot {\bf u}_a} A_a({\bf n}) E({\bf n}) d^2{\bf n} + \eta_a, 
\label{eq:voltage}
\end{equation}
where $g_a$ is the complex gain of the instrument, $A_a({\bf n})$ is the voltage primary beam in that direction, ${\bf u}_a \equiv \mathbf{x}_a/\lambda$ is the position vector of the antenna in units of wavelength,   and $\eta_a$ represents the noise for that receiver element. The measured visibility for a pair of elements $a, b$ is given by the short time-averaged complex correlation $V_{ab} = \langle \mathcal{E}_a^* \mathcal{E}_b \rangle$, which 
is then related to the sky intensity as
\begin{equation}
V_{ab}= g_a^* g_b \int d^2 \mathbf{ n} A_a^*(\mathbf{n}) A_b(\mathbf{n}) I(\mathbf{n}) e^{-i 2\pi \mathbf{n} \cdot \mathbf{u}_{ab}} + \langle \eta^*_a \eta_b \rangle,
\end{equation}
where $I(\mathbf{n)} \propto \langle |E|^2\rangle (\mathbf{n})$ is the flux density of direction $\mathbf{n}$, and the baseline vector $\mathbf{u}_{ab}=\mathbf{u}_b-\mathbf{u}_a$.  
For $a \neq b$, we expect $\langle \eta^*_a \eta_b \rangle =0$ as the noise of the two receivers are independent of each other, and by definition,  $V_{ab} = V^*_{ba}$. Thus, if the sphericity of the sky is neglected, each visibility can be regarded as a Fourier component of the  primary beam weighted sky radiation intensity. Once the primary beam, baseline vectors and complex gains are calibrated, it is then possible to synthesize all the visibilities and reconstruct the sky image, either by a direct (inverse) Fourier transform of the measured visibilities, or more commonly, by first interpolating the measured visibilities on to a regular grid (gridding), then making a Fast Fourier transform (FFT) (see e.g. \citet{Thompson_Book}). Such synthesis imaging reconstruction uses the data collected at different times and requires a large amount of computation; it is usually done offline with the stored visibility data. 

However, this approach is not well-suited for the search for or tracking of transient objects such as the fast radio bursts (FRBs). As each visibility data contains only information about one Fourier component of the observed sky, a single visibility is not particularly sensitive to the flux of a specific direction. In order to achieve sufficient sensitivity in a sky direction, one must make the image synthesis, which requires the whole visibility data set. But if we are to detect the sudden flashes, the output must be sampled on millisecond or sub-millisecond time scales. For large interferometer arrays with a large number of baselines, this would require an extremely large amount of data storage, rendering the whole system impractical.  Thus, in order to detect the transients such as the FRB, the {\em beam-forming} technique is often applied. A `beam' can be formed from the received signal of array elements so that it is sensitive to the flux density of a specific sky direction. It is then possible to sample the beam output at very short times in order to search for short bursts in that direction. The system can be designed so that once a burst is detected at a certain moment, the corresponding block of raw data is saved for more detailed analysis at later time, otherwise the huge amount of short-time sampling data of beam output can be discarded, leaving only some sort of average on longer time scales. 

An often used approach of beamforming is based on the weighted sum of the  voltages of the array elements, which we shall refer to as electric voltage beamforming (EBF). A beam formed in this way can be expressed as 
\begin{equation}
J (\mathbf{k}) = \sum_a w_a(\mathbf{k})  \mathcal{E}_a
\label{eq:sum}
\end{equation}
where $w_a(\mathbf{k})$ denotes the complex weight of antenna $a$ for beam direction $\mathbf{k}$. A simple choice of the weight is 
\begin{equation}
w_a(\mathbf{k})= \frac{1}{|g_a|} e^{-i \phi_a} e^{2\pi i \mathbf{k} \cdot \mathbf{u}_a},
\label{eq:weight}
\end{equation}
where we have denoted $g_a= |g_a| e^{i\phi_a}$, then the phases of all antennas are equal for the direction $\mathbf{n}=\mathbf{k}$.  The power of the beam is 
\begin{eqnarray}
S (\mathbf{k}) & \propto & \langle |J (\mathbf{k})|^2 \rangle, \nonumber\\
&=& \sum_{a\neq b} \int A^*_a(\mathbf{n}) A_b(\mathbf{n}) I(\mathbf{n}) e^{-2\pi i(\mathbf{n}-\mathbf{k})\cdot \mathbf{u}_{ab}}  d^2\mathbf{n} + \sum_{a} |g_a|^{-2} \langle |\eta_a|^2 \rangle
\label{eq:S_EBF}
\end{eqnarray}
where the second term is due to the thermal noise. It is also possible to apply various tapering schemes on the weight  $w_a(\mathbf{k})$ to obtain better performance on various aspects, e.g. reduced side lobe amplitudes. One can then take $S(\mathbf{k})$ as the input of the  de-dispersion and burst search program. Multiple digital beams pointing to different direction $\mathbf{k}$ can be formed at the same time, provided sufficient computing power is available. For each beam we can search for radio pulses with a range of dispersion measures.  

However, instead of forming beams from the electric voltages,  it is in principle possible to form the digital beams from the cross-correlation of the electric voltages, i.e. visibilities. We shall call this approach the cross-correlation beamforming (XBF):
\begin{equation}
S  =\sum_{a,b} \omega_{ab} V_{ab} 
\label{eq:XBF_def}
\end{equation} 
or in matrix form, $
S =\tr(\mathbf{W\cdot V})$. 
If we set the visibility weight $\omega_{ab}$ as the product of the voltage weights : $\omega_{ab} = w_a^* w_b $,
where $w_a, w_b$ are given by Eq.~(\ref{eq:weight}), and set $W_{aa}=0$, then
\begin{equation}
S(\mathbf{k})= \sum_{a \neq b} \int A^*_a(\bm{n}) A_b(\bm{n}) I(\mathbf{n}) e^{-2\pi i(\mathbf{n}-\mathbf{k})\cdot \mathbf{u}_{ab}}  d^2\mathbf{n}
\end{equation}
which reproduce the cross-correlation part of Eq.~(\ref{eq:S_EBF}). For the direction of beam, this is equivalent to the direct Fourier transform image synthesis by using only the visibility data measured at that moment. While it is not difficult to conceive this way of beam forming, the only mention of this possibility in the literature that we are aware of is that of \citet{Masui2019}, where it was introduced as ``beam-formed visibility'' and studied in the context of FFT beam forming.  

It is obvious that mathematically, any beam generated with the EBF approach can also be generated by the XBF with the appropriate weight factor. However, the reverse is not true, the XBF allows forming beams which are not possible to generate using the EBF approach. 
For example, in the XBF, we can selectively set some $\omega_{ab}$ to zero, thus masking out the contribution of these visibilities. A possible scenario in which this may be applied is for a dense array, where the adjacent elements of the array have strong cross-coupling, which may affect the observation. Masking the weights for the adjacent pairs could reduce such cross-couplings, while still retaining the other visibilities to provide sufficient sensitivity and angular resolution.

In this work we study the principle and applications of cross-correlation beam forming via simulation. The paper is organized as follows: in Sec.2, we first consider a toy model for illustration,  which is a small linear array, with a simple model of cross-couplings, in which analytical results can be obtained. We shall investigate how the cross-coupling would affect the beam-forming observations. In Sec.3, we consider the XBF, and discuss the practical issue of the hardware design for XBF, and estimate its computing load. In Sec.4,  we consider the Tianlai Cylinder Pathfinder Array \citep{Li:2020ast}, which has strong couplings for neighboring feeds on the same cylinder reflector. Finally, we summarize our results and conclude in Sec.5.

\section{Cross Coupling Effects}

We consider a perturbative linear coupling model, where the voltage at element $b$ could induce a proportionally strong voltage at element $a$.  In this model, 
\begin{equation}
\mathcal{E}_a' = \mathcal{E}_a + \sum_b \epsilon_{a,b} \mathcal{E}_b
\end{equation}
where $ \epsilon_{a,b}$ describes the self and cross couplings between the elements. In matrix form, 
\begin{eqnarray}
{\bf E'} &=& ({\bf I} + \bm{\epsilon}) \bf{E}, \nonumber\\
\mathbf{V}' &=& \langle \mathbf{E}' \mathbf{E}'^\dagger \rangle \nonumber\\
 &=&( \mathbf{I} +\bm{\epsilon}) \mathbf{V} ( \mathbf{I} +\bm{\epsilon}^\dagger)   \\
 &=& \mathbf{V} +\bm{\epsilon} \mathbf{V} + \mathbf{V}\bm{\epsilon}^\dagger +
  \bm{\epsilon} \mathbf{V} \bm{\epsilon}^\dagger
\label{eq:E'}
\end{eqnarray}
For example, the radio wave can be reflected from element $b$ to element $a$, with a possible delay represented by a  phase factor. 

The EBF in this case is
\begin{eqnarray}
S_{\rm EBF}' &=& \sum_{a,b} w_a^* w_b \langle \mathcal{E'}^*_a \mathcal{E'}_b \rangle \\
&=& \tr(\mathbf{W V'}) \nonumber\\
&=&\tr \left[\mathbf{W} ( \mathbf{I} +\bm{\epsilon}) \mathbf{V} ( \mathbf{I} +\bm{\epsilon}^\dagger)  \right]
\label{eq:S_coupling}
\end{eqnarray}

\begin{figure}
    \centering
    \includegraphics[width=0.8\linewidth]{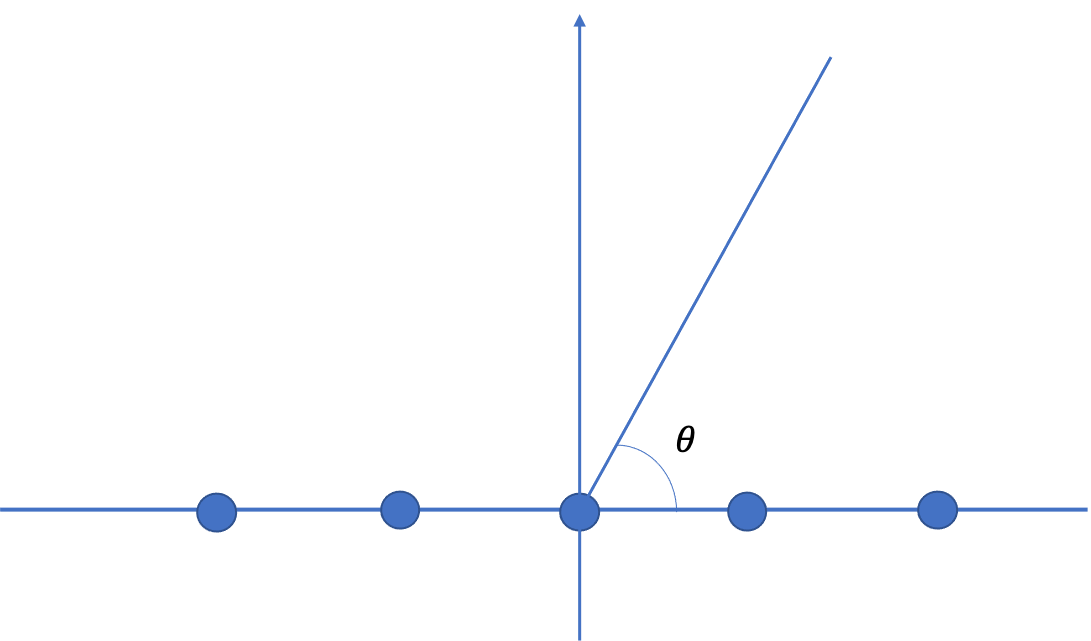}
    \caption{The 5-element linear array toy model.}
    \label{fig:toy_model}
\end{figure}

\subsection{A Toy Model: 5-element linear array}
To illustrate the effect, we first consider a toy model of a one-dimensional linear array with equally spaced 5 elements, located at $u={-2, -1, 0, 1, 2}$. We point the digital beam at the direction perpendicular to the array, i.e. the zenith, and consider the 1D beam profile on the vertical plane. A sky direction is  specified with its angle with respect to the baseline $\theta$, as shown in Fig.~\ref{fig:toy_model}. 

First we consider the case without any couplings. For simplicity we assume the antennas are ominidirectional and have a constant primary beam $A(\theta)=1$ (the expression of the EBF beam in this case is the so called ``array factor''). 
The digital beam profile of this simple array can be obtained by considering a single test source with $I(\theta') = I_0 \delta (\theta'-\theta)$, where $\theta$ is the direction of test point source, and $\mathbf{n}=(\cos\theta, \sin\theta)$.
The visibility is given by
\begin{equation}
V_{ab} =\int  I(\mathbf{n}) e^{-i2\pi \mathbf{u}_{ab} \cdot \mathbf{n}} d\mathbf{n} = I_0 e^{-i2\pi u_{ab} \cos\theta}
\end{equation}
which in matrix form is
\begin{equation}
\mathbf{V} = I_0 \left( \begin{array}{ccccc}
1 & e^{-i2\pi\cos\theta} & e^{-i4\pi\cos\theta} & e^{-i6\pi\cos\theta}& e^{-i8\pi\cos\theta}\\
e^{i2\pi\cos\theta}    &1 & e^{-i2\pi\cos\theta}  & e^{-i4\pi\cos\theta}  & e^{-i6\pi\cos\theta} \\
e^{i4\pi\cos\theta}    & e^{i2\pi\cos\theta}  &1 & e^{-i2\pi\cos\theta}  & e^{-i4\pi\cos\theta} \\
e^{i6\pi\cos\theta}    & e^{i4\pi\cos\theta}  &e^{i2\pi\cos\theta}   & 1&e^{-i2\pi\cos\theta}  \\
e^{i8\pi\cos\theta}    & e^{i6\pi\cos\theta}  &  e^{i4\pi\cos\theta}  & e^{i2\pi\cos\theta} & 1 
\end{array}\right)
\end{equation}

We consider the case where the beam is pointed to a direction $\theta_b$ with respect to the one dimensional baseline, so in the $(x,z)$ coordinates, the unit vector is $\mathbf{k}=(\cos\theta_b, \sin\theta_b)$. From Eq.~(\ref{eq:weight}), and taking the instrument complex gains to be $g_a=1$, we have 
\begin{equation}
w_a = e^{2\pi i u_a \cos\theta_b}
\end{equation}
in our case, 
\begin{equation}
\mathbf{w}(\theta_b) = \left( \begin{array}{c}
e^{-2i 2\pi  \cos\theta_b}\\
e^{-i 2\pi  \cos\theta_b}\\
1\\
e^{i 2\pi  \cos\theta_b}\\
e^{2i 2\pi  \cos\theta_b}
\end{array}\right) 
 \stackrel{\theta_b=\frac{\pi}{2}}{\resizebox{0.8cm}{5pt}{=}} \mathbf{w}_\perp= 
 \left( \begin{array}{c} 
1\\
1\\
1\\
1\\
1\\
\end{array}\right)
\end{equation}
where the second equation defines the case where the beam pointing direction is toward the zenith.

Eq.~(\ref{eq:S_EBF}) is reduced in 1D to
\begin{eqnarray}
S (\mathbf{k}) &=& \sum_{a,b}  \int d\mathbf{n}~  I(\mathbf{n}) e^{-i 2\pi [(\mathbf{n} - \mathbf{k}) \cdot \mathbf{u}_{ab}]}\\
&=& \mathbf{w}^\dagger \mathbf{V} \mathbf{w}
\end{eqnarray}
The visibility matrix $V$ is Hermitian, i.e. $\mathbf{V=V}^\dagger$, so the output $S$ is always real. 

\subsection{Cross-Coupling }
Due to the physics of such couplings, we expect that the further away the spacing between the two elements, the smaller cross-coupling, and 
we also expect $\epsilon_{a,b} \approx \epsilon_{b,a}$. As a concrete example, let us consider the following simple model. 
We assume that the coupling is only non-zero for self-coupling and the cross-coupling between the adjacent elements. 
\begin{equation}
\epsilon_{a,b}=\left\{ \begin{array}{ll}
\epsilon_0, & a=b\\
\epsilon_1, & a=b \pm 1 \\
0, & {\rm otherwise}\\
\end{array}\right.
\end{equation}
Using Eq.~(\ref{eq:E'}), we have
\begin{equation}
\mathbf{I} +\bm{\epsilon} = \left(\begin{array}{ccccc} 
1+\epsilon_0 & \epsilon_1& & & \\
\epsilon_1& 1+\epsilon_0&\epsilon_1 & & \\
& \epsilon_1& 1+\epsilon_0&\epsilon_1 & \\
& & \epsilon_1& 1+\epsilon_0&\epsilon_1 \\
& & & \epsilon_1& 1+\epsilon_0
\end{array} \right)
\end{equation}
Note that $\epsilon_0, \epsilon_1$ in general are complex numbers which varies over frequency, and $\bm{\epsilon}$ is generally not Hermitian. The $\epsilon_0$ merely changes the normalization, which after re-normalizing the beam profile it would have no effect, so for simplicity of discussion, we ignore the self coupling term and and also consider a real $\epsilon_1$. Below, we set $\epsilon_0=0, \epsilon_1=0.1$. 

\begin{figure*}
\centering
\includegraphics[width=0.99\textwidth]{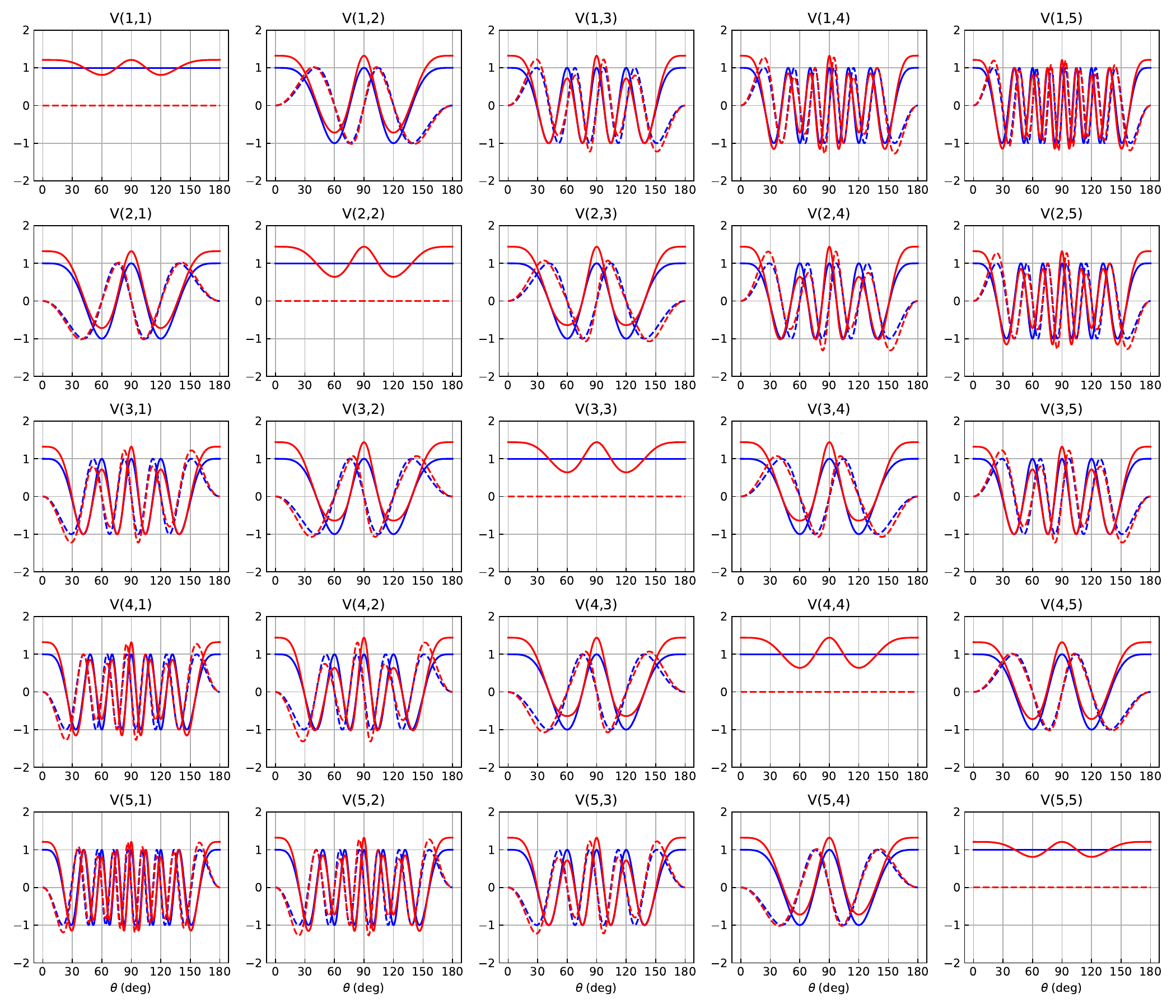}
\caption{The real (solid line) and imaginary (dash line) parts of visibilities $V_{a,b}$ as a function of the source position angle $\theta$. The blue curves show the case without couplings, and the red curves show the case with cross-couplings. }
\label{fig:V}
\end{figure*}

In Figure \ref{fig:V}, we show the visibilities, including both their real (solid line) and imaginary (dashed line) parts, as a function of the source position angle $\theta$. Here we have neglected the primary beam profile. In the real world, the antennas primary beam pattern is often limited to a much smaller range, so it is also most interesting to discuss the range of angles near the main lobe of the array.
From Eq.~(\ref{eq:E'}), we see that in the presence of cross-couplings, the observed visibility $\mathbf{V}'$ could be much different from the coupling-free case $\mathbf{V}$.  In the present case, even though the cross-couplings are only between the adjacent elements, the visibility for the non-adjacent elements is also affected, as each element in the pair also has couplings with their adjacent elements.

Although we have set $\epsilon_0=0$ in this example, the auto-correlations are also affected by the cross-couplings, which is positive near the center of the beam ($\theta \sim 90^\circ$), but oscillates over the whole range of $\theta$. This happens for all diagonal elements $V_{aa}$, but for the two elements at the end of the linear array, i.e. $V_{11}$ and $V_{55}$, the coupling-induced corrections are smaller, as they have only one neighboring element each, while for the three elements in sider, i.e. $V_{22}, V_{33}, V_{44}$, the coupling is larger. In this toy model, we also have $V_{11}=V_{55}$,  and $V_{22} = V_{33} = V_{44}$, as the coupling contributions are exactly the same in these cases. 
In the non-diagonal sub-panels of  Figure \ref{fig:V}, which corresponds to the cross-correlations, the visibility for coupling-free case and coupled case appear to be not that much different, though some deviations can be seen. 

\begin{figure*}
\centering
\includegraphics[width=0.99\textwidth]{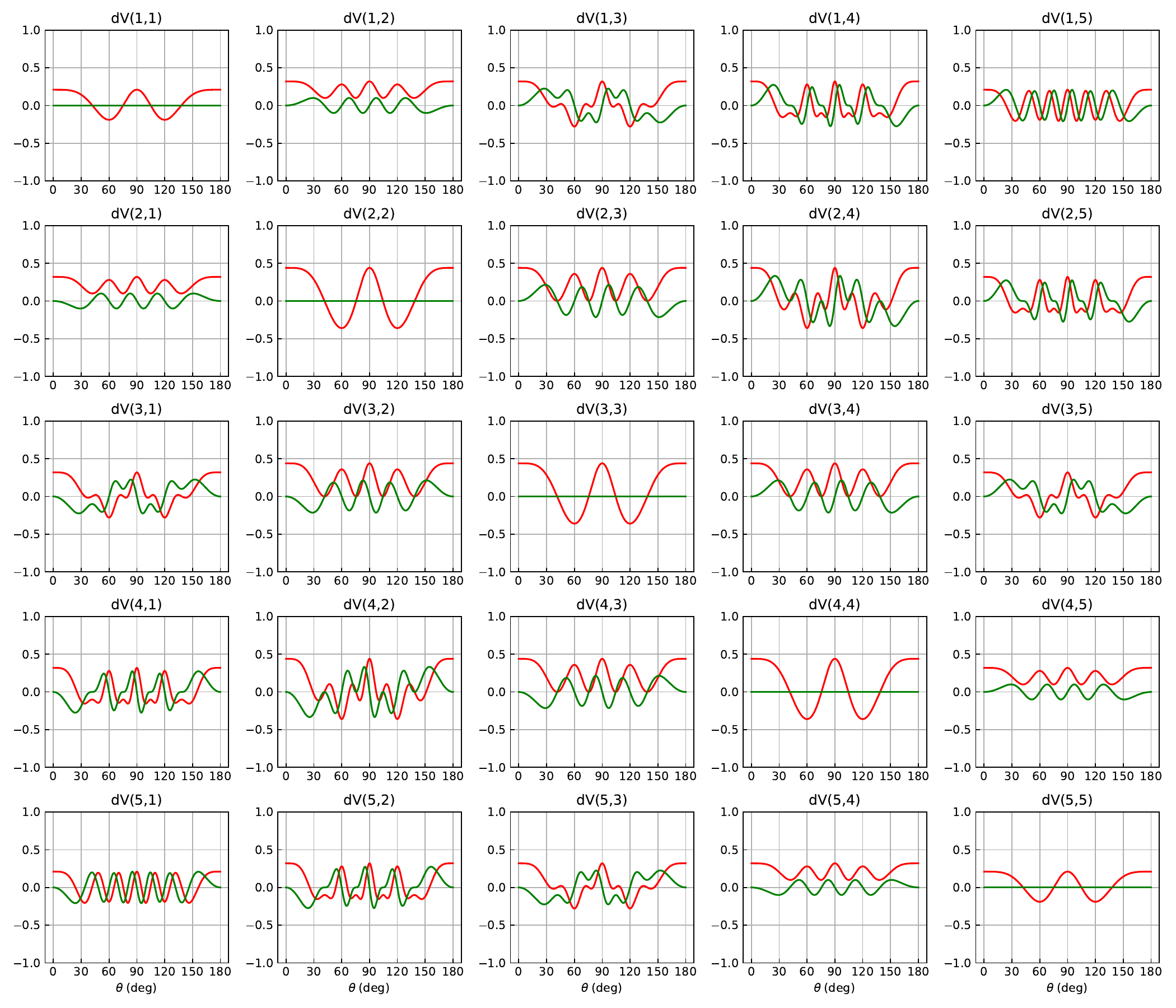}
\caption{The real (red) and imaginary (green) parts of the difference $V'_{ab}-V_{ab}$  as a function of  $\theta$ in units of degree.}
\label{fig:dV}
\end{figure*}

To see this more clearly,  in Figure \ref{fig:dV} we show the difference between the coupled case and the non-coupled case $|V'_{a,b}-V_{a,b}|$.  Except for the auto-correlation case, we see that the differences include both real and imaginary parts, and their curves have similar shapes but non-zero offsets, which indicate that the coupling induced perturbation affects the phase of the visibility. 
The difference for the diagonal elements (auto-correlation) are largest, and the cross-correlation for the adjacent elements come next, while for the non-adjacent elements are smaller, but still significant. This shows that for the non-diagonal terms,  not only the adjacent pair visibilities (such as  $V_{12}, V_{23}$, etc.) are affected, but also the non-adjacent ones $(V_{13}, V_{14}, V_{15})$. For example, although the element 1 and 5 are not adjacent with each other,  $V_{15}$ is still affected by the presence of coupling, because the leading order contribution $\Delta V_{15}$ includes terms such as $\epsilon_1 V_{14}$ and $\epsilon_1 V_{25}$, i.e. first order in $\epsilon_1$, which have comparable magnitude with the leading perturbation for adjacent pairs, though there may be less number of such perturbation terms contributing to the non-adjacent pairs. 

In the EBF approach, the beam formed is $ S' (\mathbf{k}) = \mathbf{w}_E^\dagger(\mathbf{k}) \mathbf{V}' \mathbf{w}_E(\mathbf{k})
= \tr \left[ \mathbf{W}_E \mathbf{V}'\right] $, where $\mathbf{w}_E(\mathbf{k})$ denotes the natural weighting factor given by  Eq.~(\ref{eq:weight}), and $\mathbf{W}_E =\mathbf{w}_E  \mathbf{w}_E^\dagger $. In this toy model,
\begin{eqnarray}
\mathbf{W}_E(\theta_b) &=& \left( \begin{array}{ccccc} 
1 & e^{-i2\pi\cos\theta_b} & e^{-2i2\pi\cos\theta_b} & e^{-3i2\pi\cos\theta_b} & e^{-4i2\pi\cos\theta_b} \\
e^{i2\pi\cos\theta_b}  & 1 & e^{-i2\pi\cos\theta_b} & e^{-2i2\pi\cos\theta_b} & e^{-3i2\pi\cos\theta_b}  \\
e^{2i2\pi\cos\theta_b} & e^{i2\pi\cos\theta_b} & 1 & e^{-i2\pi\cos\theta_b} & e^{-2i2\pi\cos\theta_b}   \\
e^{3i2\pi\cos\theta_b} &  e^{2i2\pi\cos\theta_b} & e^{i2\pi\cos\theta_b} & 1 & e^{-i2\pi\cos\theta_b} \\
e^{4i2\pi\cos\theta_b} & e^{3i2\pi\cos\theta_b} &  e^{2i2\pi\cos\theta_b} & e^{i2\pi\cos\theta_b} & 1  
\end{array}\right)
\label{eq:WE1}
\end{eqnarray}
which for the $\theta_b=\frac{\pi}{2}$ case, is reduced to 
\begin{equation}
\mathbf{W}_{E\perp} = \left( \begin{array}{ccccc} 
1 & 1 & 1 & 1 & 1 \\
1 & 1 & 1 & 1 & 1 \\
1 & 1 & 1 & 1 & 1 \\
1 & 1 & 1 & 1 & 1 \\
1 & 1 & 1 & 1 & 1 
\end{array}\right)
\label{eq:WE2}
\end{equation}

\begin{figure*}
\centering
\includegraphics[width=0.49\textwidth]{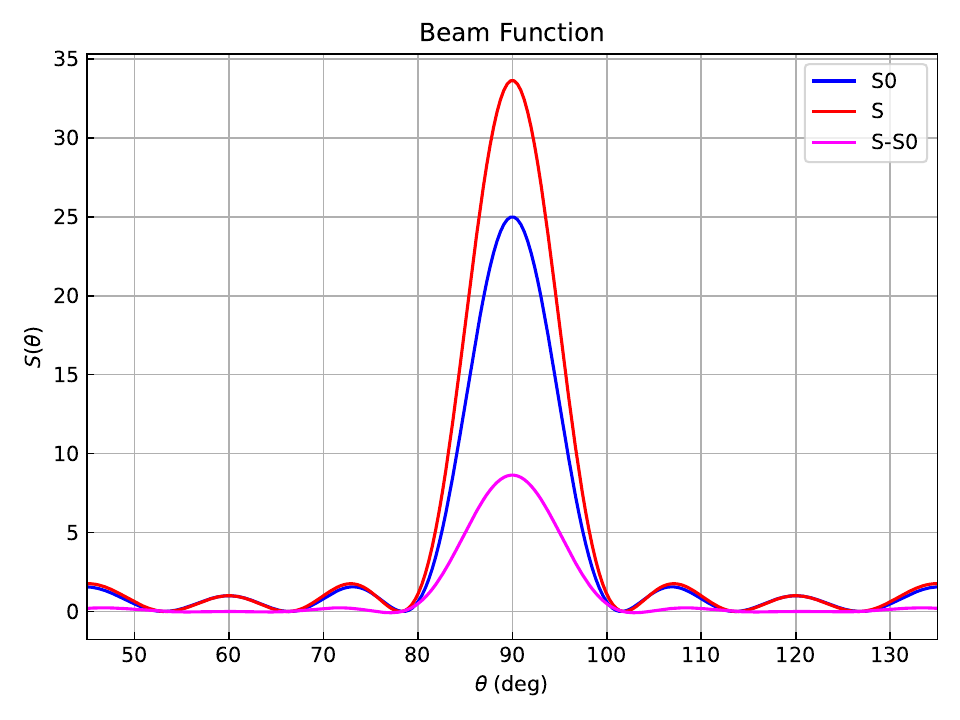}
\includegraphics[width=0.49\textwidth]{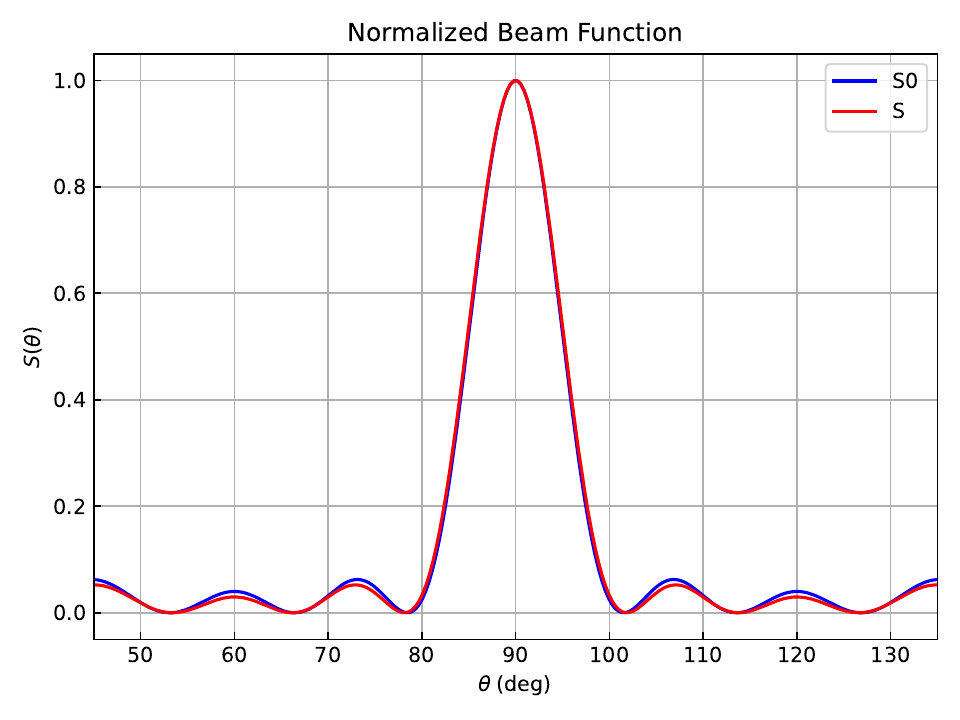}
\caption{Left: Beam profile for the coupling-free case (blue) and coupled case (red) and their difference (magenta). Right: Normalized beam profile for the coupling-free case (blue) and coupled case (red), with $S(90^\circ)=1.$ }
\label{fig:S}
\end{figure*}

In the left panel of Figure \ref{fig:S}, we plot the profile of  the EBF beam without cross coupling $S_0(\theta)$, and the EBF beam with cross-couplings $S_{\rm E} (\theta)$, and their difference in the range of $45^\circ < \theta < 135^\circ$. There are more peaks for larger range of angles, but the angles will be limited by the antenna primary beam in most real world applications. When the coupling is introduced, there is substantial change in the amplitude of the peak. But in this case, the two beams have very similar shapes, and the nulls nodes and side lobe peaks are located at the same position as the unperturbed case. Because of this, if we normalize the beam to the peak of the primary beam so that $S(90^\circ)=1$, then as shown in the right panel of Figure \ref{fig:S}, the two beam profiles are almost identical. So in this model, the cross-couplings do not strongly affect the normalized beam profile.

\section{Cross-Correlation Beam Forming (XBF)}
As noted in the Introduction, we can form beams by using the visibilities. We now illustrate this using the toy model introduced above. 

\subsection{The Toy Model Case}

According to the definition Eq.~(\ref{eq:XBF_def}),  
\begin{equation}
S_X = \sum_{a,b} \omega_{ab} V'_{a,b} = \tr \left[ \mathbf{W} \mathbf{V}'\right],
\end{equation}
but now we have more freedom in choosing $\mathbf{W}_X$. 
Again we consider the toy model discussed in the last section. We can form exactly the same beam by employing Eqs.(\ref{eq:WE1})-(\ref{eq:WE2}).
However, we have more general possibilities. For example, we could have a  $\mathbf{W}_X$ of the following form:
\begin{equation}
\mathbf{W}_d = \left(\begin{array}{ccccc}
0 & 1 & 1 & 1 & 1\\
1 & 0 & 1  &  1 & 1\\
1 & 1 & 0 & 1& 1 \\
1 & 1 & 1& 0 & 1\\
1 & 1 & 1 & 1 & 0
\end{array}\right)
\label{eq:W_d}
\end{equation}
Only the cross-coupling is used in the beam forming, while the auto-correlations, which usually have a non-zero background is avoid. 

\begin{figure}
\centering
\includegraphics[width=0.45\textwidth]{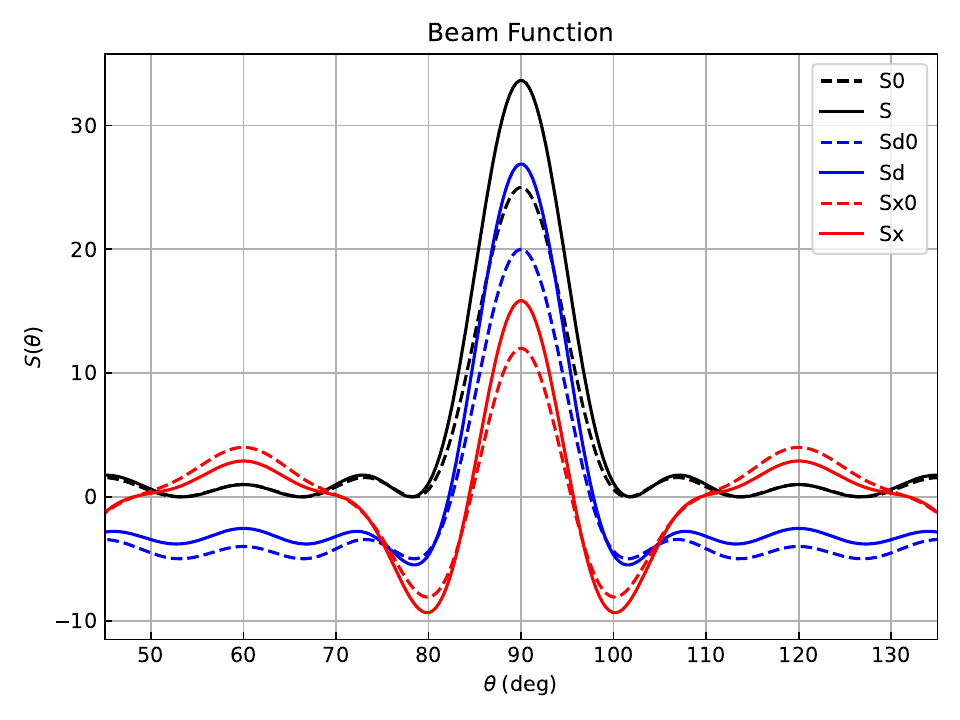}
\includegraphics[width=0.45\textwidth]{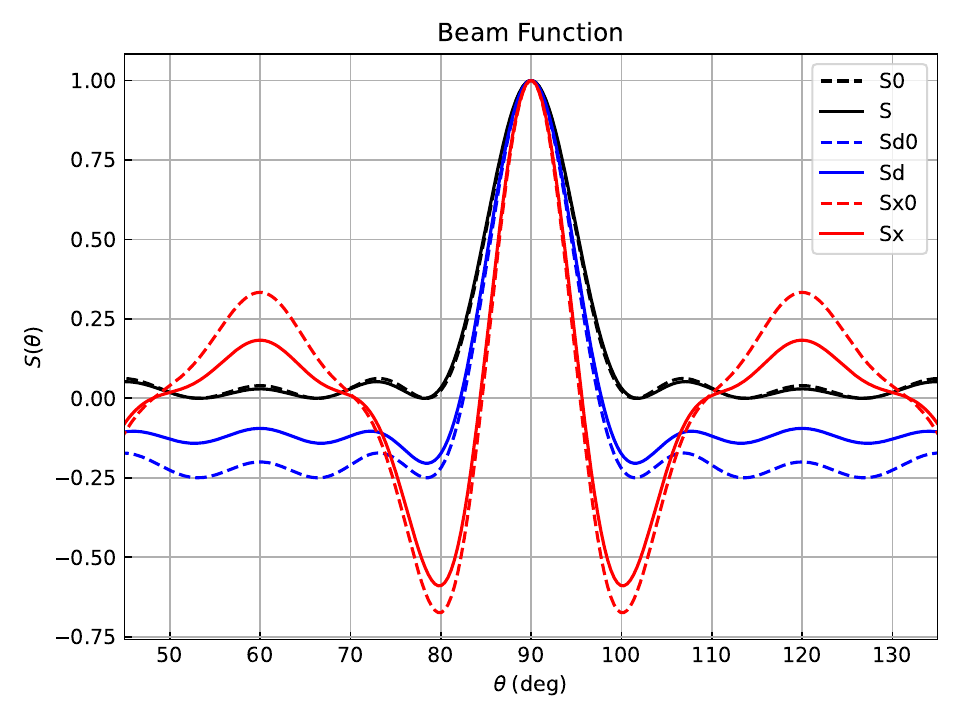}
\caption{The unnormalized (left panel) and normalized (right) XBF beam profiles.  The coupling-free case ($S_0$, dashed line) and coupled case ($S$, solid line), with the weight matrix $\mathbf{W}_d$ (Eq.\ref{eq:W_d}, blue) and $\mathbf{W}_x$ (Eq.\ref{eq:W_X}, red). For comparison, the corresponding EBF beam profiles (black) are also plotted.}
\label{fig:Sdx}
\end{figure}

Additionally, considering the cross-couplings, we may also remove the  adjacent element pairs which is affected most. In the toy model case, this will be 
\begin{equation}
\mathbf{W}_X = \left(\begin{array}{ccccc}
0 & 0 & 1 & 1 & 1\\
0 & 0 & 0  &  1 & 1\\
1 & 0 & 0 & 0 & 1 \\
1 & 1 & 0 & 0 & 0\\
1 & 1 & 1 & 0 & 0
\end{array}\right)
\label{eq:W_X}
\end{equation}
In this particular case, as too many elements are set to zero, the effect may not be ideal. However, in the more realistic case, it could be interesting. We will try this out in the next section, with the Tianlai case.

In Figure \ref{fig:Sdx} we plot the profile of the beams formed with $\mathbf{W}_d$ and $\mathbf{W}_X$, the left panel shows the unnormalized plot, and the right panel shows the profiles all normalized to the same main peak height. The cases without and with cross-couplings are marked as $S_{d0}, S_d$ (blue lines) and $S_x, S_{x0}$ (red lines) respectively. For comparison, we also plot the EBF beams $S_0, S$ (black lines). Note that with XBF beams, the beam can become negative outside the main lobe, this is a price for removing the auto-correlation. The auto-correlation is  usually more noisy than the cross-correlations, as the averaged noise on the cross-correlation is zero, while for auto-correlation it is the system temperature. Thus, one might want to remove the auto-correlation to reduce noise, but then instead of nulls we get the negative beam.  The amplitude of the XBF beams are also lower, as fewer visibilities are used. This is itself not a problem, for the signal can always be amplified to suitable range, and the system will be calibrated, so the measured flux is not affected. Compared with removing only the diagonal elements (produced with $\mathbf{W}_d$), the beam which also removes the contribution of adjacent elements (produced with $\mathbf{W}_X$),  has even lower amplitude at the peak without normalization, and its minimum is at a deeper negative, or after normalization, a deeper negative trough. It also has more pronounced side lobes. This is due to the removal of baselines of adjacent pairs, which contribute to the power on larger angular scales. Furthermore, we see that there are still differences in the beam with couplings and without couplings, because even though the visibilities of adjacent neighbors are not used in the XBF, the couplings still affect the result indirectly as contributions to non-adjacent pair voltages.

These cases show that with XBF, there are indeed more possibilities of beamforming, including some which are not possible in EBF. However, at least for the above toy model, these new beam forms do not exhibit clear advantages over the simple EBF beam.

\subsection{Amount of Computation}
Besides the characters of the beams, one also needs to consider the practicality and cost, if the XBF is to be implemented in a real case. For interferometer arrays, the beamforming needs to be performed at nearly real time on site. 
In principle, if one can store a large amount of radio frequency(RF) data, the beamforming can also be processed afterwards, but for an interferometer array which has a large number of receiver feeds, doing this would produce a huge amount of data, which is impractical except for very short durations. This is especially true for the EBF case, where the beamforming (weighted sum) operation requires using the RF voltage before any integration. For the XBF, the cross-correlations (visibilities) have to be computed at real time, but once this is done, the phase information of the wave is preserved in the visibility data, and the final beamforming part (weighted sum of the visibility) can in principle be processed later. However, the number of visibilities is proportional to the square of the number of receiver feeds, so for arrays with large number of feeds, the amount of data is huge, even if the amount of the data can be compressed somewhat by integrating the visibilities for a short time. Thus, although the XBF would allow post-processing in principle, in practice it still needs to be processed at nearly real time.

Now we consider the amount of computation in the beamforming. We assume that the array uses the FX architecture, i.e. the voltage of each array element is first digitized, and make a Fast Fourier transform (FFT) of fixed length to transform the data to frequency domain, then making the cross correlations or beamforming. 

We can estimate the amount of computation as follows: suppose we have $N_a$ antennas, and plan to form $N_b$ digital beams. The FFT length is assumed to have $N_f $ frequencies, and each integration time contains $N_t$ FFT intervals (data frames). The amount of computation for each FFT of length $N_f$ requires $\sim  N_f \log N_f$ computing operations, and as there $N_a$ antennas, each FFT interval requires $N_a (N_f \log N_f)$ computation for the array. This part is however common  for the EBF and XBF methods. Next we consider the correlation, EBF and XBF.

{\bf Correlation}: there are $N_f$ frequencies, at each frequency, for the $N_a$ antenna, there are $N_a (N_a+1)/2$ pairs (including auto-correlation), so the computation during one FFT interval is $\frac{N_a (N_a+1)}{2} N_f$, and computation of correlations during one integration time is 
$\frac{N_a(N_a+1)}{2} N_f N_t$. \\

{\bf EBF}:  during one integration time,  there are $N_b$ beams, each antenna each frequency multiply one weight factor, and adding the results together, so the total is  $N_b (2N_a) N_f N_t$.\\

{\bf XBF}:  during one integration,   there are $N_b$ beams, at each frequency, compute the beam, with the visibilities assumed already known. This  can be computed once per integration time, so the total is : 
$N_b 2\frac{N_a(N_a+1)}{2} N_f$.\\

So the amount of XFB vs. EBF is 
\begin{equation}
\frac{\mathrm{XBF}}{\mathrm{EBF}} \sim \frac{ N_b N_a(N_a+1) N_f}{N_b (2N_a) N_f N_t} \sim \frac{N_a}{2N_t}.
\end{equation}
Thus, the larger the array element number $N_a$, and the shorter integration number $N_t$, the EBF is more economic compared with the XBF.

As an example, for the Tianlai array (see next section) we have 
$$N_f=2048,$$ 
$$T_{\rm FFT} = \frac{N_f}{\rm Sampling~ Rate}=2048/250 \MHz= 8.192 \mu s.$$
$$t_{\rm int} = 0.1 \mathrm{ms},$$ 
$$N_t=0.1 \mathrm{ms} /8.192 \mu s= 12.2.$$
$$N_a =128 $$

Thus, $$\frac{\mathrm{XBF}}{\mathrm{EBF}} \sim 10$$

We see that in this case, the XBF method does require more computing power to realize. If we increase the integration time to e.g. 1ms, then the two could have approximately the same amount of computation. However, as FRBs typically have millisecond or sub-millisecond pulses, it is desirable to keep the integration time short so that the full pulse profile can be captured. Thus, although the XBF are generally more flexible and may offer certain advantages, in terms of computing resource required it would be more expensive. Note also that the above estimation is based on the theoretical amount of computational operations, but in the real world, the efficiency is not 100\% and varies a lot, depending largely on the particular hardware configuration. Besides the computing speed itself, a specific CPU or GPU also has limited data communication bandwidth and size of the memory, etc., these generally do not match each other perfectly for all applications. Thus, the actual required resources may also not scale up exactly by the ratio estimated above.

\section{The Tianlai Cylinder}
In this section we consider the beam forming for the Tianlai cylinder experiment. We first introduce the Tianlai cylinder array briefly, and construct a simple model which can describe the primary beam of the cylinder, and the cross-coupling between the feeds, then investigate the beamforming of the Tianlai cylinder array with both the EBF and XBF method, and present the results. 

\subsection{A brief introduction of the Tianlai cylinder array}
The Tianlai experiment consists of a dish pathfinder array \citep{Wu2021dish} and a cylinder pathfinder array \citep{Li:2020ast,Zuo:2018}, which aim to test key technologies for cosmological surveys using the 21-cm intensity mapping method \citep{Chen2012,Xu_2014}. The cylinder pathfinder array consists of three reflectors, marked as A, B, and C cylinders, and along the focus line of  each cylinder is a uniformly spaced linear array of feeds, with the same maximum distance between the two ends of the array (12.4 meters), but with slightly different number of feeds on each cylinder (31, 32, 33 respectively), to suppress grating lobes \citep{Zhang_2016}. The feeds are designated as A1, A2, ..., A31, B1, B2, ...B32, C1, C2,... C33, an interferometric baseline can be specified by the designation of the pair of feeds. The basic parameters and configuration of the cylinder array are summarized in Table \ref{tab:cylinder_properties}, see \citet{Li:2020ast} for more details. Initially, these arrays were equipped with a correlator that provided visibilities with a second-level sampling rate \citep{Niu2019}, which is sufficient for sky map reconstruction \citep{ZUO2021100439}. However, the Tianlai arrays have a relatively wide field of view, which is suitable for blind search of FRB with high speed digital backend equipped, and now EBF backends have been installed on both the dish array \citep{Yu:2022RAA} and the cylinder array \citep{Yu:2024bqq}.

\begin{table}[h!]
    \centering
    \caption{Basic parameters of the Tianlai Cylinder Pathfinder Array}
    \begin{tabular}{c c}
    \hline
         Parameters & Values \\
    \hline 
         Number of cylinders  & 3            \\
         Reflector E-W diameter & 15 m              \\
         Reflector N-S length & 40 m           \\
         Number of receivers  & 96       \\
         f/D & 0.32                      \\
         Average SEFD & 67.18 kJy        \\
         Latitude & 44.15\degree N       \\
         Longitude & 91.80\degree E      \\
         Current Observing frequency & 685--810 MHz          \\
    \hline
    \end{tabular}
    \label{tab:cylinder_properties}
\end{table}

The 96 dual linear polarization receiver feeds and the subsequent analog front end generates a total of 192 electric voltage outputs, which are then digitized. Each of these digital signal are  transformed to the frequency domain by Fast Fourier Transform (FFT). The signals are then fed through a network switch to a number of processing units, each unit are fed the data of the same block of frequencies, but from all different receivers, for the purpose of the beam forming.  This beam-forming part of processing units, called the B-machine, forms the beams according to the EBF method we outlined above. 
The beamformed stream data is then processed by the transient search pipeline, as described in detail in  \citet{Yu:2024bqq}. The current Tianlai cylinder B-engine consists 12 GPU nodes, each GPU node has two servers and each server is equipped with an NVIDIA RTX 3060 and 8 GB VRAM. These are used to form 96 beams from 96 feed inputs. An upgrade for more receivers and beams are being considered.

\subsection{Primary Beam and Cross-Coupling Model}

We model the primary beam of the Tianlai as \citet{Tianlai2023}:
\begin{equation*}
    A(\hat{\vb*{n}}) = A_{\text{NS}} (\sin^{-1}(\hat{\vb*{n}} \cdot \hat{\vb*{x}}); \theta_{\text{NS}}) \times A_{\text{EW}} (\sin^{-1}(\hat{\vb*{n}} \cdot \hat{\vb*{y}}), F, \theta_{\text{EW}}),
\end{equation*}
where $\hat{\vb*{x}}$ and $\hat{\vb*{y}}$ are the unit vector pointing East and North, respectively, and
\begin{equation}
    A_{\text{NS}}(\theta; \nu) = \exp[-4 \ln{2} \left( \frac{\theta}{\theta_{\text{NS}}(\nu)} \right)^2],
    \label{eq:beam-ns}
\end{equation}
\begin{equation}
     A_{\text{EW}}(\theta;F, \theta_{\text{EW}})
     \propto \int^{\frac{1}{4F}}_{-\frac{1}{4F}} A_D\left(2 \tan^{-1} u ;\theta_\text{EW}\right) e^{- i\frac{4\pi F W u}{\lambda} \sin \theta} \dd u,
    \label{eq:beam-ew}
\end{equation}
with $\theta_{\text{NS}}(\nu) = \alpha \frac{\lambda}{D_{\text{NS}}}$, in which $D_{\text{NS}} = 0.3\,\mathrm{m}$ as the size of the Tianlai cylinder feeds, $W$ is the width of the cylindrical reflector, and $A_D$ is taken as
\begin{equation*}
    A_{D}(\theta; \theta_{\text{FWHM}}) = \exp \left[ -\frac{\ln 2}{2} \frac{\tan^2\theta}{\tan^2(\theta_{\text{FWHM}}/2)} \right].
\end{equation*}
The fitted parameters used here are $\alpha = 1.04$, $F = 0.2$, $\theta_{\text{EW}} = 2.74$ \citep{Tianlai2023}.

\begin{figure*}[h!]
    \centering
    \includegraphics[width=0.99\textwidth]{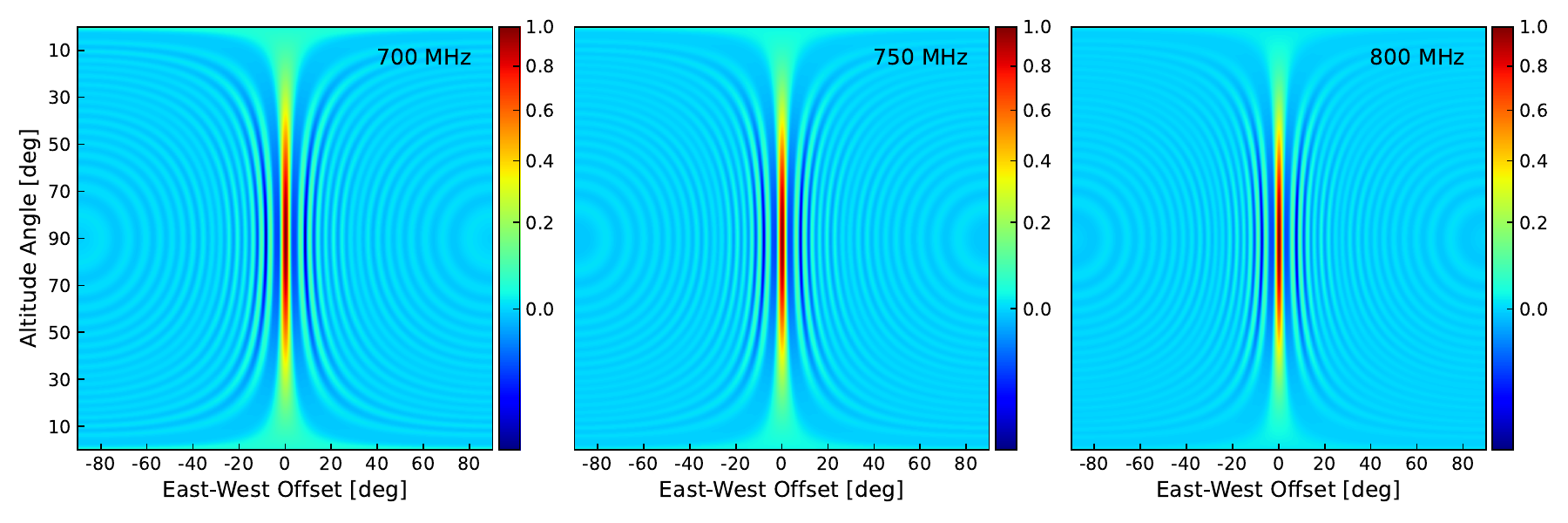}
    \caption{The 2D primary beam of Tianlai cylinder array at 700, 750 and 800 MHz.}
    \label{fig:pribeam2D}
\end{figure*}
\begin{figure*}
    \centering
    \includegraphics[width=0.8\textwidth]{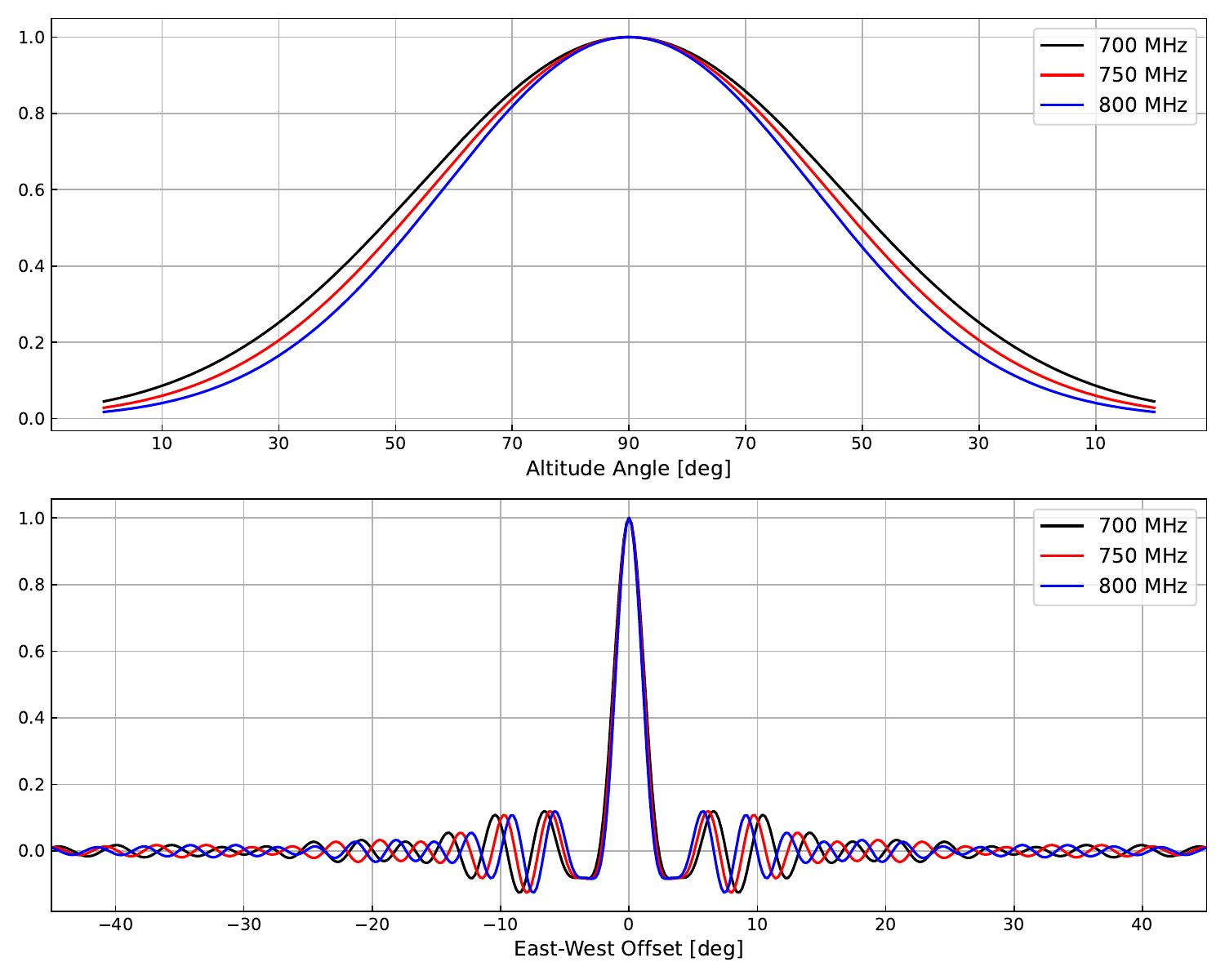}
    \caption{The 1D primary beam in the North-South direction (top) and East-West direction (bottom) of Tianlai cylinder array at 700, 750 and 800 MHz. }
    \label{fig:pribeam1D}
\end{figure*}

In Figure \ref{fig:pribeam2D}, we show the above two-dimensional primary beam model profile for the frequencies of  700, 750 and 800 MHz. In this model, the profile can be decomposed into the product of the East-West profile and the North-South profile. Due to the variation of frequency, there are slightly differences for the three frequencies, but the general shapes are still similar. 

In Figure \ref{fig:pribeam1D}, we show the primary beam profile in the East-West direction and North-South direction separately. The N-S profile are broad with a FWHM width of nearly $100^\circ$, while along the E-W direction there is a peak, with a width of about $1.5^\circ$, and multiple side lobes. In the 1D plot, the difference between different frequencies can be more clearly seen. 

\begin{figure}
    \centering
    \includegraphics[width=0.6\textwidth]{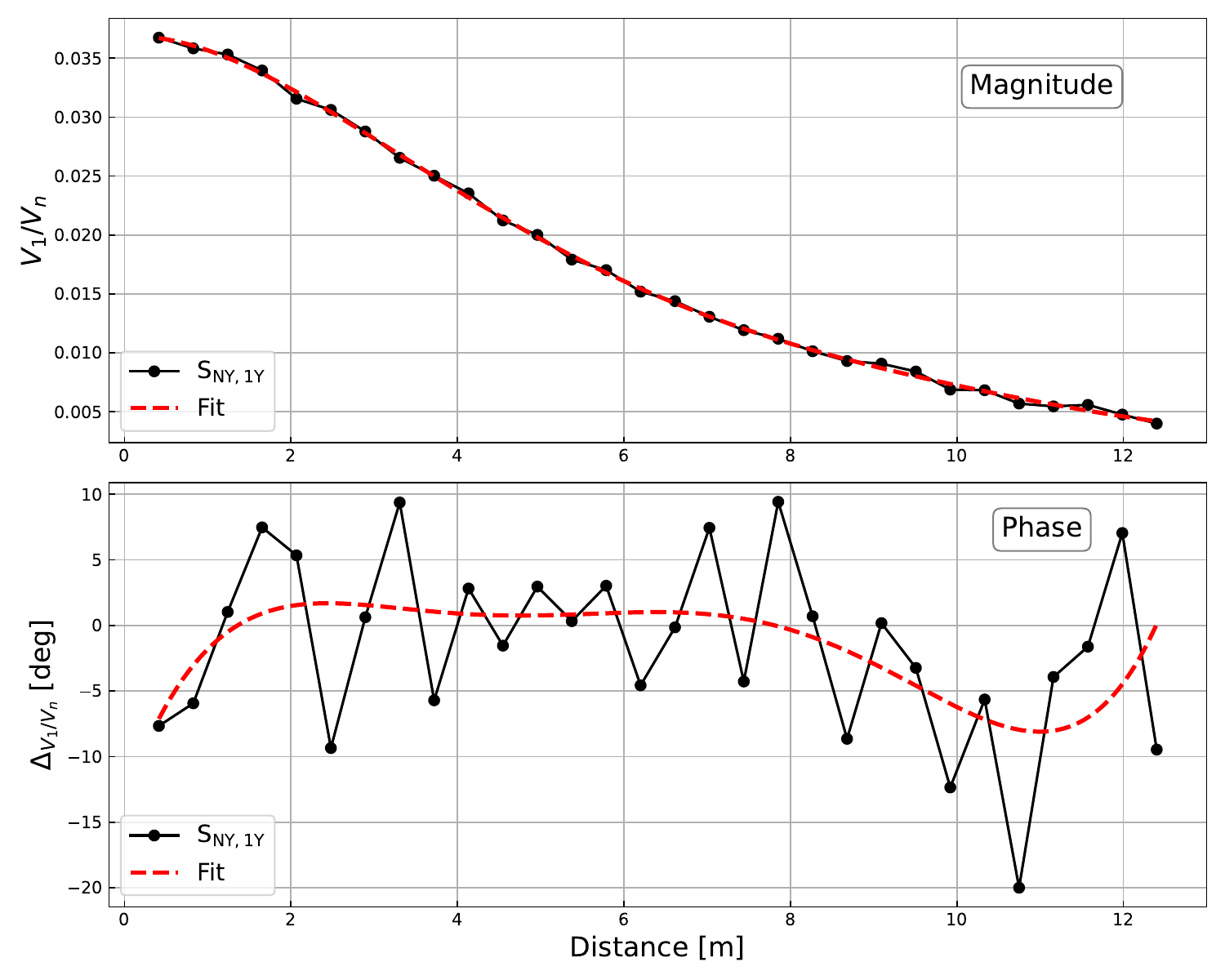}
    \caption{The fitted polynomials for cross coupling magnitude (top) and phase (bottom).}
    \label{fig:cross_coupling}
\end{figure}
\begin{figure}[h!]
    \centering
    \includegraphics[width=0.9\textwidth]{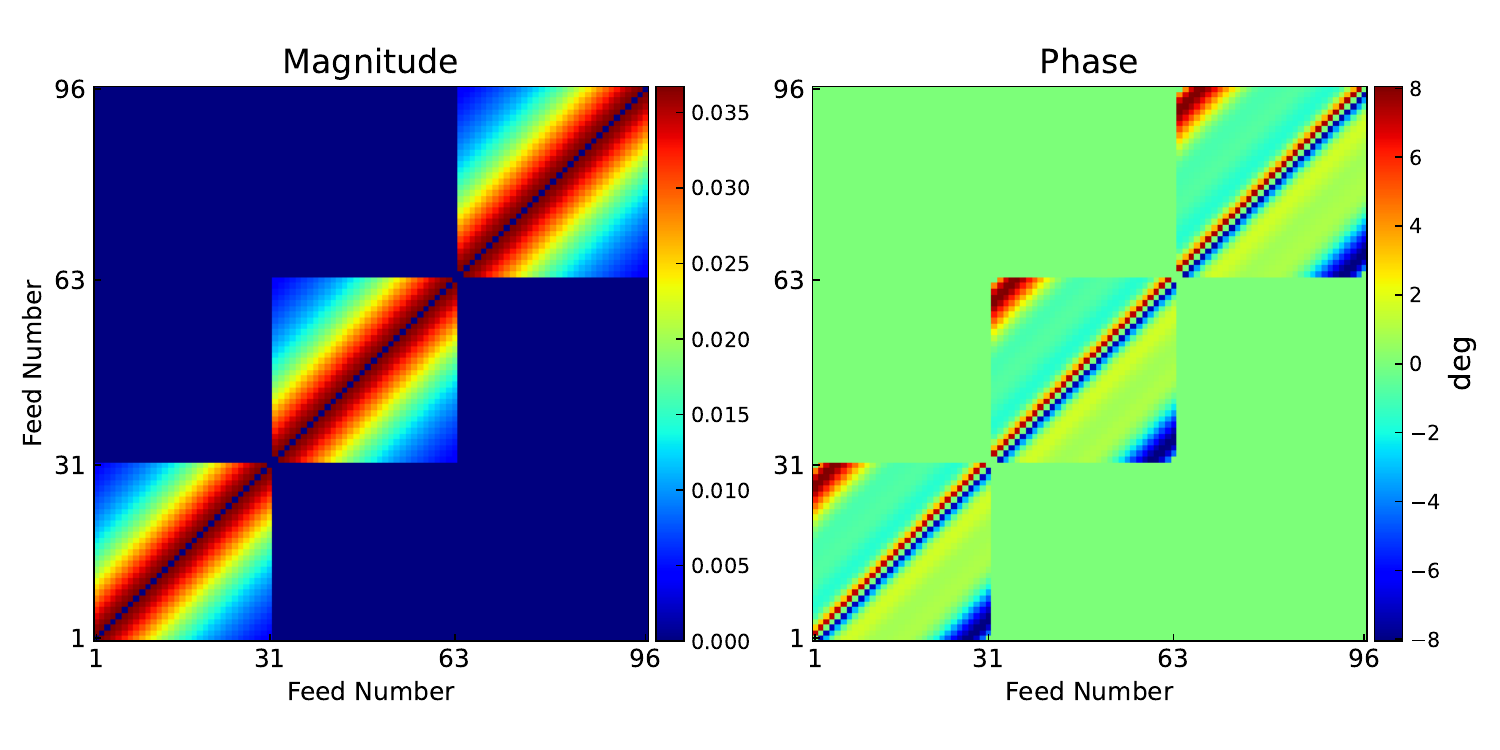}
    \caption{The adopted cross coupling coefficients}
    \label{fig:Eij}
\end{figure}

From observational data, we could see some indications of couplings between the different feeds \citep{Li_2021}, especially for the feeds located on the same cylinder reflector. This is easy to understand: not only the feeds on the same cylinder are closer to each other and are directly exposed to each other, but the main lobe of the beam is also directed to the cylinder reflector, and thus within the main lobe of the neighboring feeds on the same cylinder. On the other hand, the different cylinders are farther away from each other, and their main beams are not on the same reflector, so the cross-couplings are much weaker. 

Below, we shall neglect the cross-couplings  between different cylinders, but consider the couplings on a single cylinder.  We have performed an electromagnetic field simulation of the array, which gives such couplings coefficients \citep{Sun_2022}. The coupling coefficients actually vary with frequency, but at each frequency we can all see the same pattern that it is largest for adjacent feeds, and drops quickly when we go to the further away feeds. In Figure \ref{fig:cross_coupling}, which is based on the simulation data in \citet{Sun_2022} we can see this trend clearly. Here, the transmission coefficient $\epsilon_{a,b}$ is averaged over a frequency range and given in dBs, but drops quickly as we go from the nearest to farther away element pairs. There is also some variation in the phase, though generally much smaller than the geometric phase between the feeds. The coupling is different for the two polarizations, but similar trends can be seen.  

For demonstration we shall consider the Y-polarization, which can be modeled as
\begin{equation}
\epsilon_{i,j} = A(d_{ij}) \, e^{i\phi(d_{ij})}
\end{equation}
where $d_{ij}$ is the distance between feed $i$ and feed $j$, $A$ and $\phi$ are polynomials fitted from the amplitude and phase of the voltage ratio between feed i and j on cylinder A. This is shown in Figure \ref{fig:Eij} for both the magnitude (left panel) and phase (right panel) of the adopted cross coupling coefficients for all pairs of elements.  We can see there are three blocks clearly shown up in the matrix. These corresponds to the cross-couplings of the same-cylinder feeds. The cross-couplings across cylinders are much weaker, indeed they were ignored in \citet{Sun_2022}. Along the same cylinder, the feeds are not only located near each other, but they ``see'' each other via the reflection of the same cylinder. While there may still be some cross-coupling across the cylinders, the strength of such couplings are orders of magnitude lower. 

\begin{figure}
\centering    \includegraphics[width=0.9\textwidth]{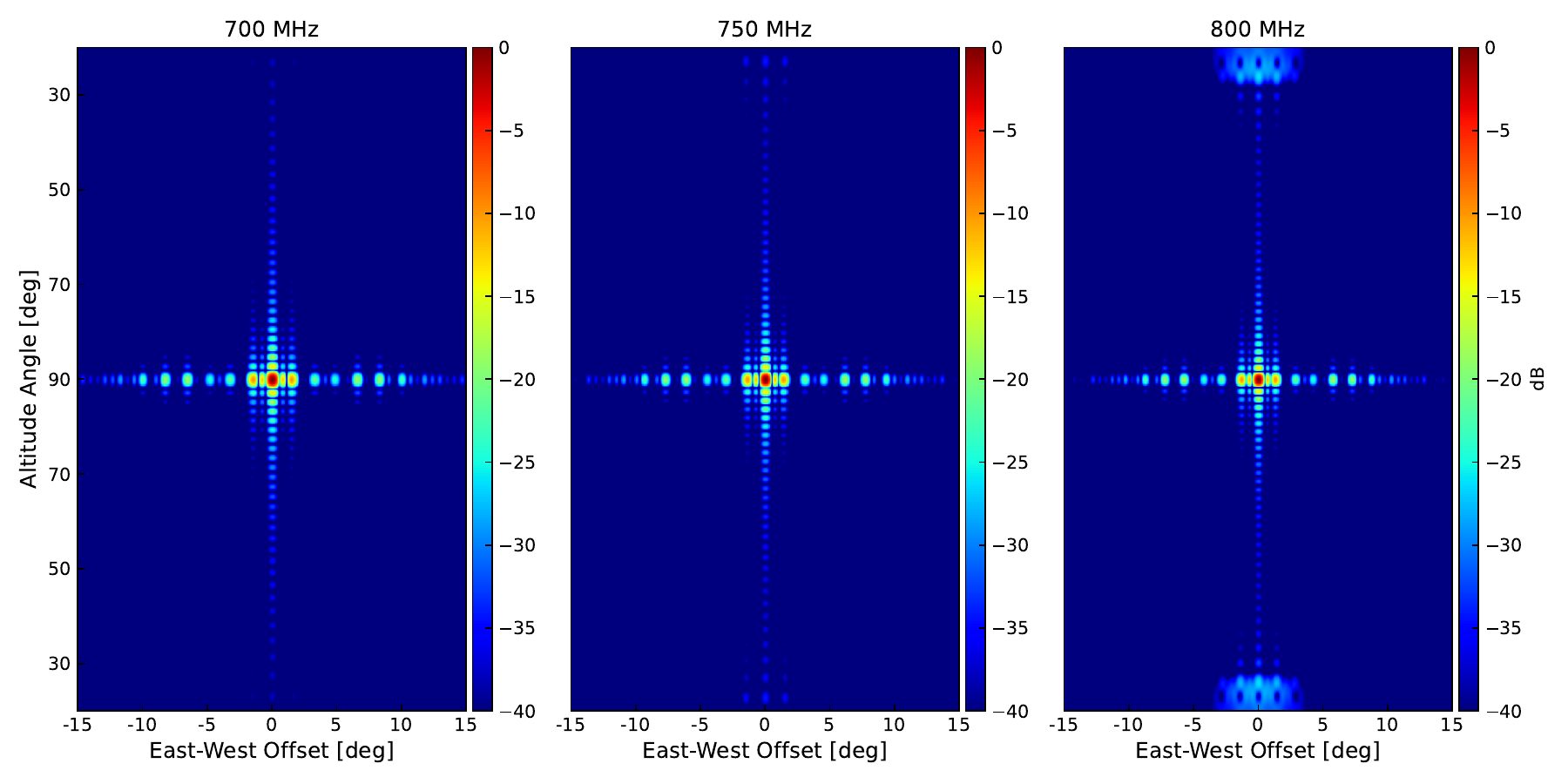}
\includegraphics[width=0.9\textwidth]{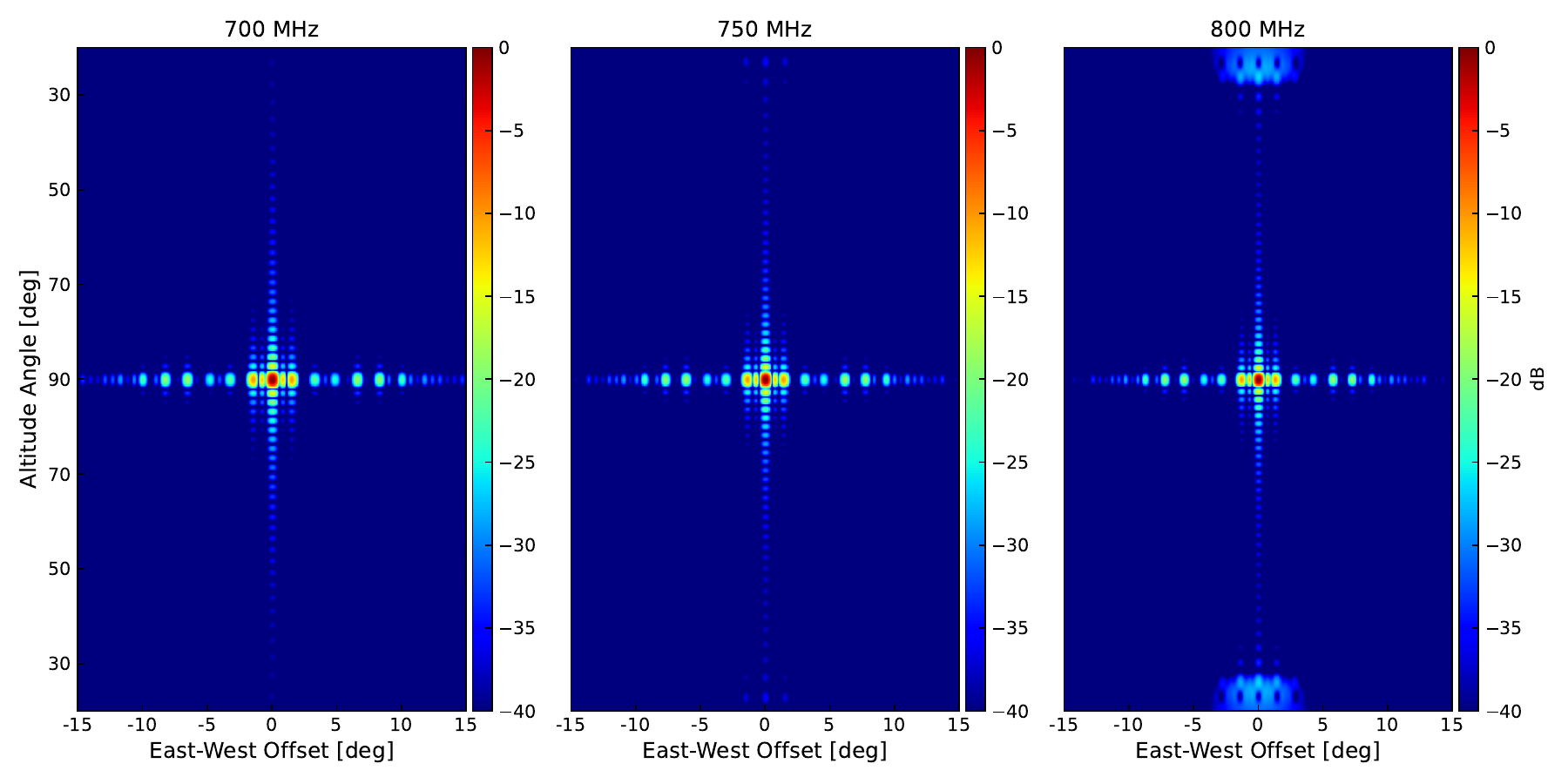}
\caption{The EBF beam (in dB) at zenith in the coupling-free case (top) and coupled case (bottom). }
\label{fig:ebf_th0_phi0}
\end{figure}

\subsection{BeamForming}

Based on the above models, we simulate the beam in the Tianlai case. The EBF beam pointing toward the zenith is shown in Figure \ref{fig:ebf_th0_phi0}, the coupling-free case is shown in the top panels, and the same case with cross-couplings is shown in the bottom panels. The three columns are for the observational frequency of 700, 750, and 800 MHz, respectively. The synthesized beams can be well decomposed into a profile along the N-S direction and a profile along the E-W direction. We can see the main lobe at zenith, and the side lobes along the E-W and the N-S directions. In the case of 800 MHz, near the edge of the plot, the side lobes of the primary beam are also visible.  However, as in the case of the primary beam, the difference between the different frequencies is not that obvious in this kind of plot, and the difference between coupling-free case and coupled case is also not very apparent.

\begin{figure}
\centering
\includegraphics[width=0.49\textwidth]{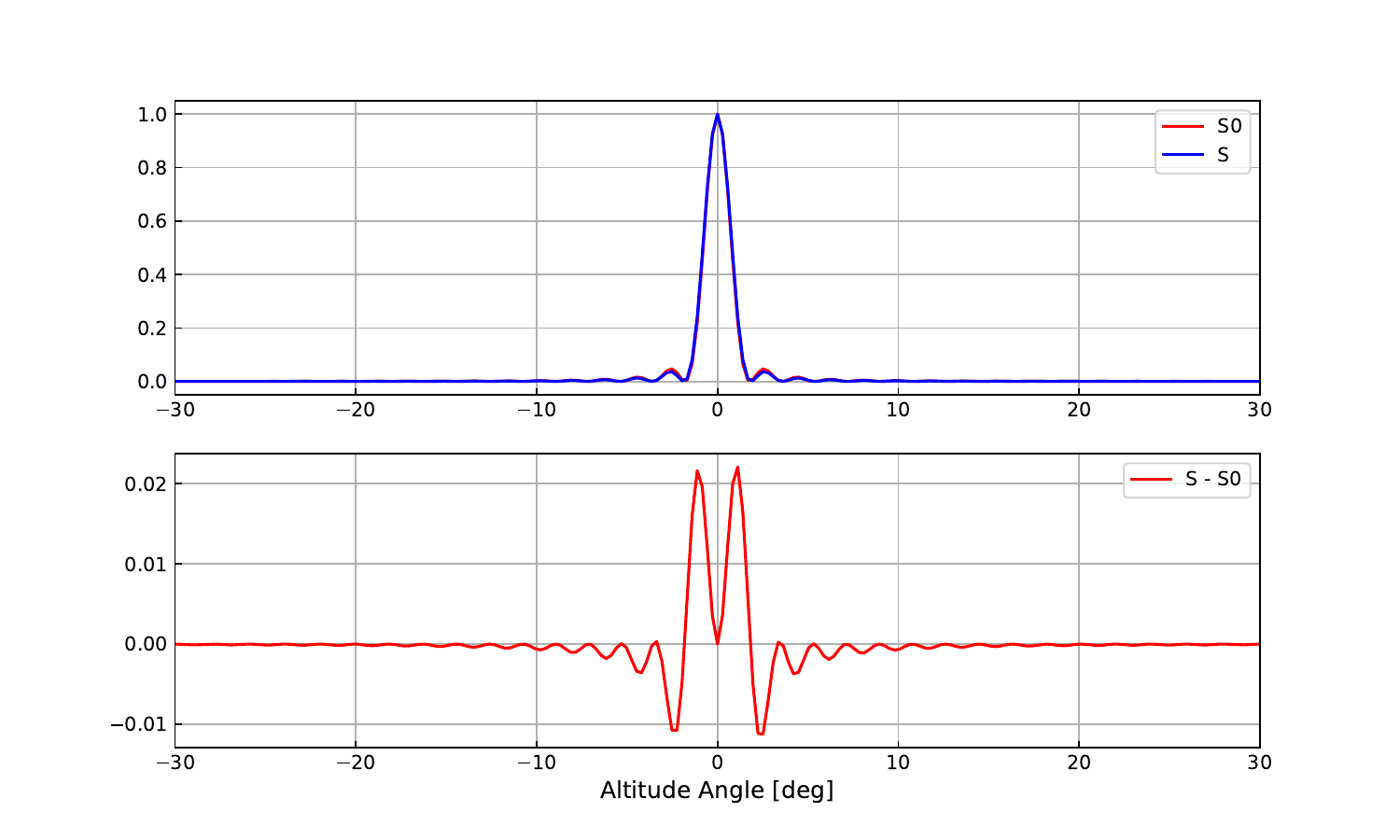}
\includegraphics[width=0.49\textwidth]{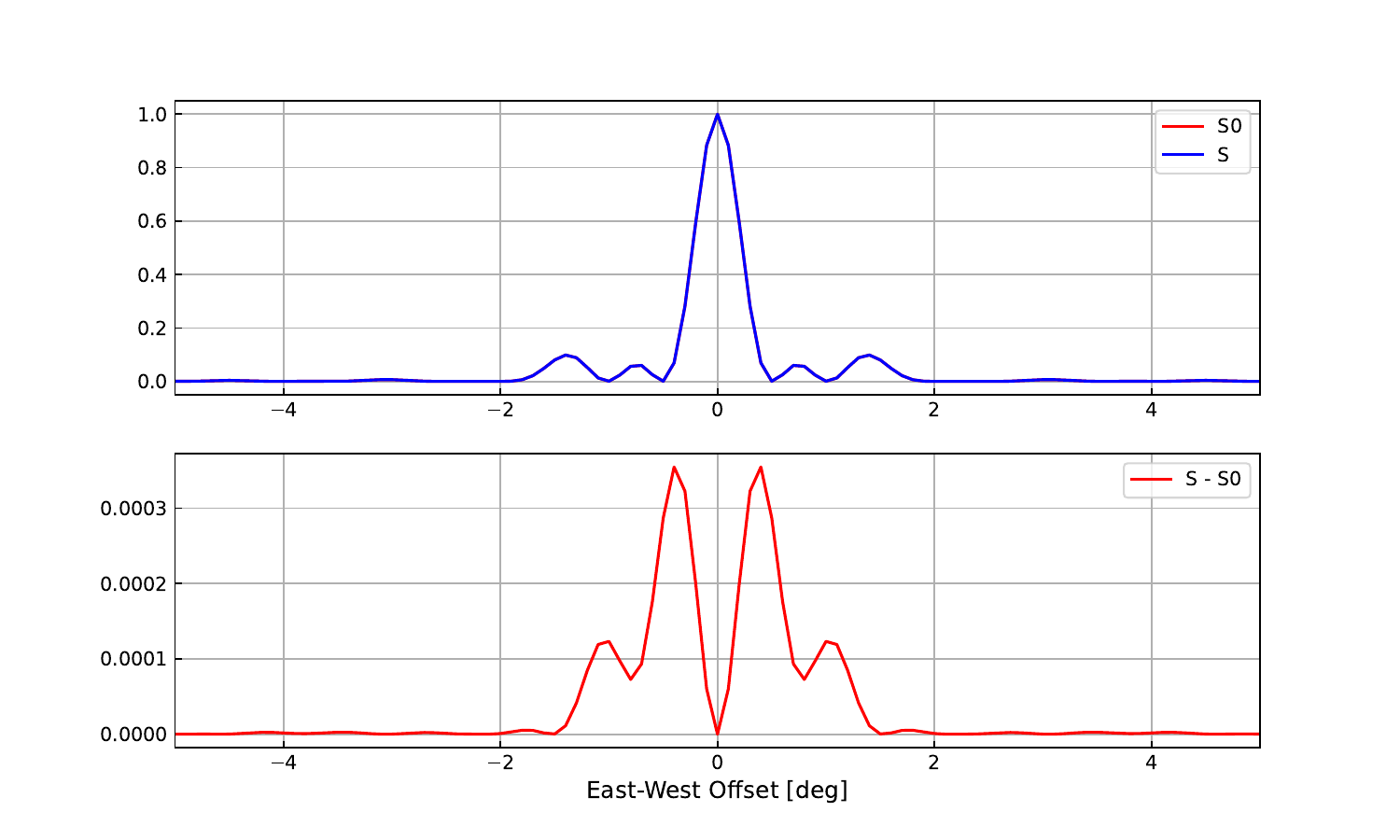}
\caption{The EBF beam 1D profiles in the North-South direction (left top) and East-West direction (right top) of the Tianlai cylinder array at 750 MHz. $S_0$ denotes the coupling-free case, $S$ denotes the coupled case. The difference in profiles $S-S_0$ are shown in the bottom panels (N-S for left bottom and E-W for right bottom).}
\label{fig:TL1D}
\end{figure}

\begin{figure}
\centering
\includegraphics[width=0.49\textwidth]{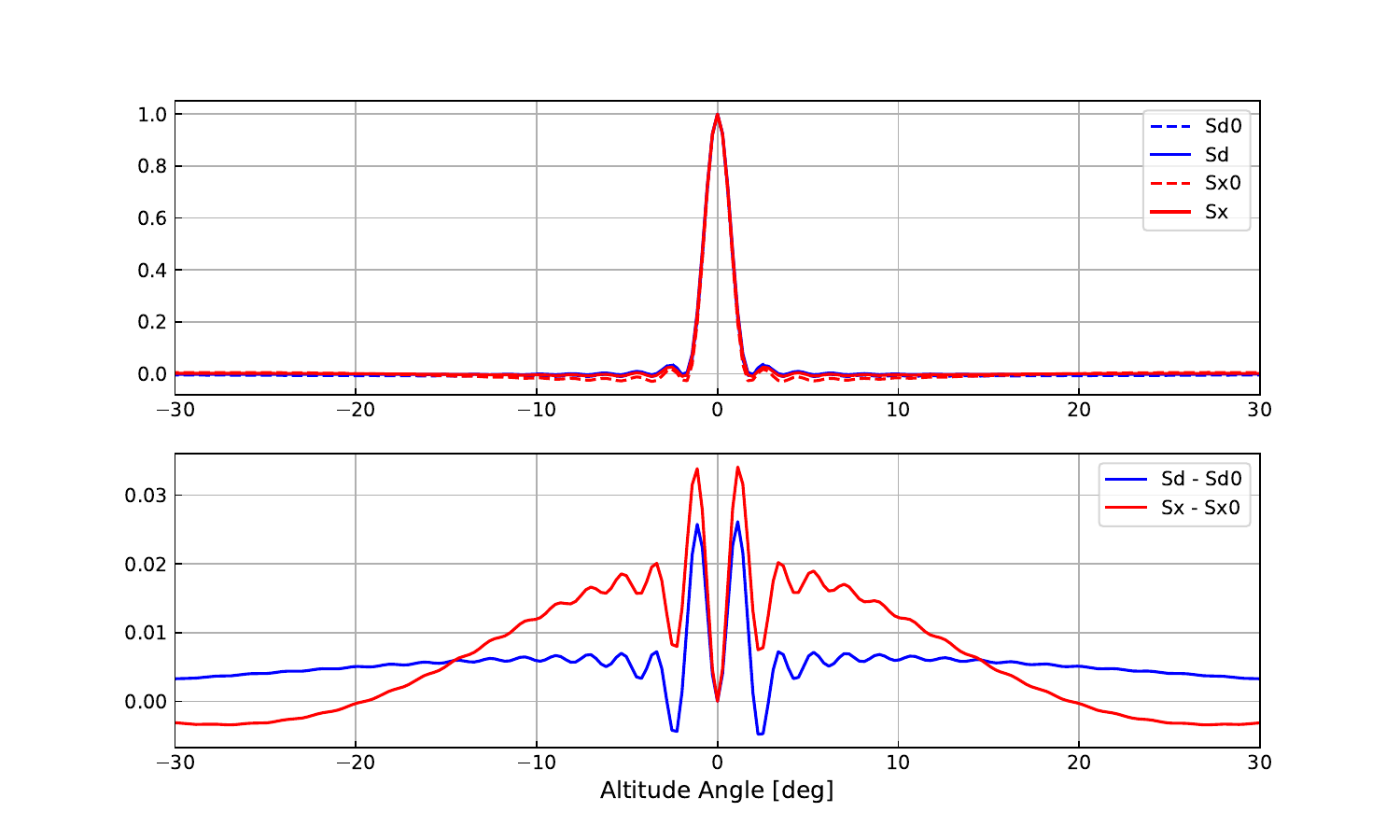}
\includegraphics[width=0.49\textwidth]{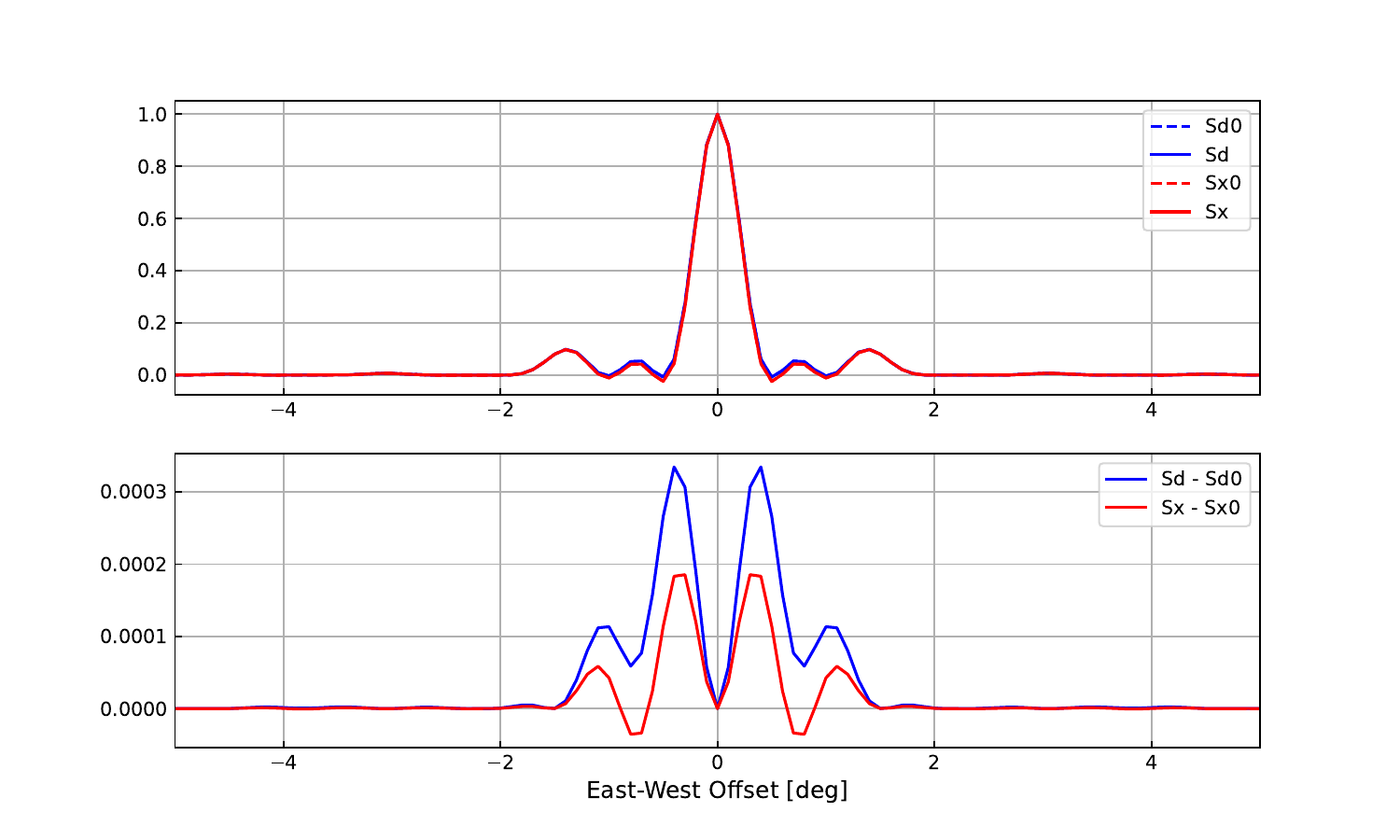}
\caption{The normalized 1D profile for the XBF beam in the North-South direction (left top) and East-West direction (right top) of the Tianlai cylinder array at 750 MHz. We have plotted the XBF formed with $W_d$ and $W_x$ as given in Eqs.~(\ref{eq:W_d})-(\ref{eq:W_X}). The differences $S_d-S_{d0}$ and $S_x-S_{x0}$ are plotted in the bottom panels.
}
\label{fig:TL1XBFDlim}
\end{figure}

In Fig.~\ref{fig:TL1D}, we plot the normalized 1D profile of the EBF beam along the N-S (left) and E-W direction (right), for both the coupling-free case ($S_0$) and coupled case ($S$). As the difference between $S_0$ and $S$ is very small, we also show the difference $S-S_0$ in the bottom panels. In both the N-S and E-W direction, the synthesized beams are modulated by the primary beam profile on larger scales, and since the primary beam of the cylinder is flat along the cylinder direction (N-S direction), we have plotted the N-S direction for a wide angular range, but of course the main lobe of the synthesized beam is much smaller than the primary beam in both the N-S and E-W directions. There are side lobes along both directions, but in the E-W direction, there are only three sets of receivers, so the synthesized beam has multiple side lobes, some of which can be seen in this figure, but the side lobes are attenuated by the primary beam profile, so the ones at larger angles are not apparent in this plot.

As shown in these figures, the cross-coupling changes the pattern of the beam profile. However, in the normalized beam profile, the differences between the coupling-free case (red line) and the coupled case (blue line) are quite small. The curves almost coincides with each other, the differences are barely visible at the peaks and troughs in the profile. To see the differences more clearly, in the bottom panels we show the difference $S-S_0$. We see that for the N-S direction the maximum differences are about $2\%$, while for the E-W direction they are even smaller, only at the $10^{-4}$ level. We have met similar situation in the toy model discussed earlier--despite the cross-couplings, the variation in the normalized beam profiles at a specific frequency is not that large. Here the cross-coupling is more complicated, but still does not strongly alter the shape of the beam. 

We also plot the normalized 1D profile of the XBF beam, where the auto-correlations and the cross-correlations between adjacent feeds are removed in Figure~\ref{fig:TL1XBFDlim}. In this plot. we show the case without couplings (dashed line) and with couplings (solid line). For comparison, we also plotted the EBF lines. Again, we see that the normalized beam profile is only slightly affected by the cross-couplings. While in principle the XBF suffers less from the cross-couplings, due to the removal of the most affected adjacent receivers, the difference is actually more apparent due to the lack of the auto-correlation and short baselines. Thus, in the present case, the EBF could still work very well in the presence of cross-couplings, and at least from the perspective of beam profile, the XBF does not offer a clear advantage.

\section{Summary}
In this paper, we have studied two methods of beamforming, which produce the radio intensity of specific sky directions, which are well-suited for the search for transient sources. The EBF method has been widely used, which is based on the weighted sum of the voltages of different array elements with appropriate phases, as defined in  Eqs.~(\ref{eq:sum})-(\ref{eq:S_EBF}). Alternatively, one can also use the XBF method, which is based on the weighted sum of visibilities, i.e. the short time cross-correlation of the voltages of array elements, as defined in Eq.~(\ref{eq:XBF_def}). 

A possible motivation for using the XBF is that it is more flexible, as it can in principle form beams which can not be formed by the EBF method. For example, The XBF can be used to form beams without including the contribution from adjacent feed receivers, which may have strong cross-couplings. We investigated the impact of cross-couplings on beamforming by first considering a simple toy model of 5-element linear array. However,  we find that although the cross-coupling could have significant impact on the visibilities, the beam may indeed vary over frequency, at a single frequency, the normalized beam profile changes very little. Also, even with the removal of auto-correlations and cross-correlations of adjacent feeds in the XBF beam, the effect of the cross-couplings is not completely removed. A more apparent change in the beam profile is induced by the lack of short spacing baselines rather than the reduction of cross-couplings. These changes may make the beam become negative at some points, and increase the amplitude of the side lobes. 

We have also applied the beam-forming simulation to the Tianlai cylinder pathfinder array. Presently, there is an EBF beamformer on the Tianlai cylinder array, its observations were reported in \citet{Yu:2024bqq}. The investigation of the EBF vs. XBF methods can be useful in choosing a design for an upgrade of the Tianlai beam former, which is being actively considered. Based on an electromagnetic field simulation, we construct a model of this cylinder array, with primary beam profile and cross-couplings between elements modeled by fitting the simulation result. These couplings decrease with increasing distance between the feeds. We then simulated with the cross-couplings set to zero or the model value. We find that like the toy model case, the difference between the coupling-free case and coupled case in the normalized beam profile is very small. Again, one can remove the cross-correlation of adjacent feeds in the XBF approach, but the beam profile changes more due to the lack of short baselines. 

An important consideration in the practical implementation of the beam-forming scheme is the cost of computing. As noted above, the XBF method is more flexible, it can also produce the equivalent of the usual beam formed with the EBF method, as well as other forms of beam profile. However, it often requires more computing resources for large arrays, as the computation of the EBF scales with the number of elements of the array $N_a$, while for XBF this scales with the number of cross-correlations which is $O(N_a^2)$. This makes the XBF less desirable. 
In the Tianlai case, the amount of computations required by the XBF beamforming is about 10 times that of the EBF, but the advantage is not obvious. Based on these results, it seems that the EBF should be the preferred method of beamforming in the Tianlai case.

\begin{acknowledgements}
We thank Ms. Meiting Zhao, Ms. Qiuxiang Fan and Dr. Lin Shu of the Institute of Automation, Chinese Academy of Science for discussion on the Tianlai beamformer hardware configurations.
The Tianlai Cylinder Pathfinder Array Transient Digital Backend is built with the funding from the Repair and Procurement Grant of the Chinese Academy of Science. For this research we acknowledge the support of the National SKA program of China ( Nos. 2022SKA0110100 and 2022SKA0110101), the National Natural Science Foundation of China (Nos. 1236114814, 12203061, 12273070, 12303004).
\end{acknowledgements}

\bibliographystyle{raa}
\bibliography{ms}

\end{document}